# What can we learn about the lipid vesicle structure from the small-angle neutron scattering experiment?
## (Investigation DMPC vesicle structure by small angle neutron scattering)

M.A. Kiselev[1], E.V. Zemlyanaya[2], V.K. Aswal[3], R.H.H. Neubert[4]

[1] Frank Laboratory of Neutron Physics, Joint Institute for Nuclear Research, Dubna 141980, Moscow region, Russia

[2] Laboratory of Information Technologies, Joint Institute for Nuclear Research, Dubna 141980, Moscow region, Russia

[3] Solid State Physics Division, Bhabha Atomic Research Centre, Trombay, Mumbai 400 085, India

[4] Institute of Pharmaceutical Technology and Biopharmacy, Martin-Luther-University, Halle (Saale) D-06120, Germany

Correspondence to:
M.A. Kiselev
Frank Laboratory of Neutron Physics
Joint Institute for Nuclear Research
Dubna 141980, Moscow region
Russia
tel. 7-096-21-66977
fax. 7-096-21-65484
E-mail: Kiselev@jinr.ru

**Abstract**

Small angle neutron scattering (SANS) on the unilamellar vesicle populations (diameter 500 Å and 1000 Å) was used to characterize lipid vesicles from dimyristoylphosphatidylcholine (DMPC) at three phases (gel, ripple, and liquid). Parameters of vesicle populations and internal structure of the DMPC bilayer were characterized on the basis of the Separated Form Factor (SFF) model. Vesicle shape changes from about spherical in the liquid phase to elliptical in the ripple and gel ones. It is true for vesicles prepared via extrusion through pores with the diameter 500 Å. Parameters of the internal bilayer structure (thickness of the membrane and the hydrophobic core, hydration, and surface area of lipid molecule) were determined on the basis of the Hydrophobic-Hydrophilic (HH) approximation of neutron scattering length density across the bilayer $\rho(x)$ and of the Step Function (SF) approximation of $\rho(x)$. It was demonstrated in frame of HH approximation that DMPC membrane thickness in liquid phase ($T = 30^{o}$C) depends on the membrane curvature. Prepared via extrusion through 500 Å diameter pores, vesicle population has the following characteristics: average radius of 275.6±0.5 Å, polydispersity 27%, thickness of membrane 47.8±0.2 Å and of hydrophobic core 20.5±0.3 Å, surface area 61.0±0.4 Å$^2$ and number of water molecules 11.9±0.3 per DMPC molecule. Vesicles prepared via extrusion through pores with the diameter 1000 Å have polydispersity of 48 %, and a membrane thickness of 45.6±0.2 Å. SF approximation was used to describe the DMPC membrane structure in gel ($T = 10$ $^{o}$C) and ripple ($T = 20$ $^{o}$C) phases. DMPC vesicles prepared via extrusion through 1000 Å pores were found to have membrane thickness of 49.6±0.5 Å in the gel phase and 48.3±0.6 Å in the ripple phase. The dependence of the DMPC membrane thickness on temperature was restored from the SANS experiment.

**Introduction**

Phospholipids are the main components of cell membranes. Research in the structure of phospholipids is important from a viewpoint of structural biology and biochemistry. Unilamellar vesicles are especially interesting because most of biological membranes are unilamellar. Function and properties of integral membrane proteins depend on the lipid bilayer structure. On the other hand, unilamellar vesicles are also used as delivery agents for drugs. Knowledge of their structure in nanoscale region is important for pharmacology (Nagayasu et al. 1999, Cevc et al. 2002).

Dynamic and static light scattering are commonly used to characterize the form and size of vesicles. However, these methods have the limitation of obtaining information about the thickness and the internal structure of the membrane bilayer (Pencer et al. 2001, Jin et al. 1999). Main part of modern knowledge about internal structure of phospholipids in liquid and gel phases was obtained by X-ray diffraction on the giant multilamellar vesicles, which have negligibly small membrane curvature (Nagle and Tristram-Nagle 2000). In many works SANS have been used to characterize the bilayer structure of the unilamellar vesicles at high excess of water. Membrane thickness can be found from experimentally measured radius of gyration of the bilayer by using Guinier approximation (Feigin and Svergun 1987; Knoll et al. 1981; Gordeliy et al. 1993; Balgavy et al. 1998). Calculations of bilayer parameters from the radius of gyration are based on the part of SANS curve in the interval of scattering vector $q$ from 0.03Å$^{-1}$ to 0.14Å$^{-1}$. This approach was used to calculate the bilayer thickness and lipid surface area of the DMPC membrane in liquid phase on the base of Strip Function model of the neutron scattering length density (Kucerka et al. 2004). The electron density profile or the one of neutron scattering length density is calculated by Fourier transformation from diffraction peaks intensities in the diffraction experiment on the multilamellar vesicles or oriented dry membranes (Wiener and White 1991; Nagle and Tristram-Nagle 2000; Tristram-Nagle et al. 2002). Accuracy of the structure determination depends on the space resolution of the scattering experiment

$$\Delta x \approx \frac{\pi}{q_m}, \qquad (1)$$

where $q_m$ is a maximum value of measured scattering vector. In diffraction experiment

$$\frac{\pi}{q_m} \approx 0.5 \cdot \frac{d_u}{h_m}, \qquad (2)$$

where $d_u$ is repeat membrane thickness and $h_m$ is maximum diffraction order (Kiselev et al. 2005). In access water, the DMPC membrane has $d_u = 62.7$ Å at $T = 30^{\circ}$C and $\Delta x = 4.7$ Å for the $h_m = 8$. In the SANS experiment, the coherent scattering intensity of vesicle population can be measured to $q_m \approx 0.3$ Å$^{-1}$. This gives a resolution $\Delta x = 10.5$ Å of the Fourier transformation, which limits an application of the indirect Fourier transformation for the evaluation of the internal membrane structure from SANS experiment (Glatter 1977; Glatter 1980). Design of appropriate scattering models and approximations of the scattering length density across the bilayer $\rho(x)$ (based on some preliminary knowledge about bilayer structure) could improve the space resolution in the evaluation of the internal membrane structure from SANS experiment. The model of randomly oriented planar bilayer was applied for identifying of the membrane thickness and the internal membrane structure (Pencer and Hallet 2000; Kucerka et al. 2004). Step Function (SF) approximation of $\rho(x)$ was applied for the investigation of oligolamellar vesicles (Schmiedel et al. 2001). This approach could be used to obtain additional information about membrane repeat distance and percentage of nonunilamellar vesicles.

Other important information about vesicle population concerns average vesicle radius and polydispersity. Hollow Sphere (HS) model was applied for the calculation of vesicle radius, vesicle polydispersity, membrane thickness and internal membrane structure (Kiselev et al. 2001; Balgavy et al. 2001). The application of HS model gives possibility to describe the internal membrane structure as two or three regions with constant scattering length density via SF approximation of the $\rho(x)$. HS model has two principal drawbacks: a) it can be used only for spherical shape of vesicles, b) $\rho(x)$ can be described only as step function, whereas neutron diffraction experiments demonstrate that $\rho(x)$ has a more complex and smooth shape (Wiener and White 1991; Gordeliy and Kiselev 1995).

A model of Separated Form Factors (SFF) allows simulating of $\rho(x)$, using almost any functions, which significantly expands the possibilities of studying the internal membrane structure (Kiselev et al. 2002). Calculated within it parameters were: the average vesicle radius <R>, the relative standard deviation of radius (polydispersity) $\sigma$, the thicknesses of membrane $d$, of hydrophobic part of bilayer $D$, and of hydrophilic part $(D-d)/2$, as well as the number of water molecules within a bilayer per DMPC molecule $n_w$. (Zemlyanaya and Kiselev 2002; Zemlyanaya et al. 2005). The parameters of DMPC vesicle form at $T = 30$ $^{\circ}$C do not depend sufficiently on the different functions used for modeling $\rho(x)$: <R> = $275 \pm 0.4$ Å and $\sigma = 27\%$ (Zemlyanaya and Kiselev 2002; Zemlyanaya et al. 2005), <R> = $272 \pm 0.4$ Å and $\sigma = 27\%$ (Kiselev et al. 2004). However, the values of $d$, $D$, and $n_w$ are sensitive to the type of function $\rho(x)$ (Kiselev et al. 2004, Kucerka et al, 2004).

In the case of homogeneous approximation, $\rho(x) =$ const, D$_2$O penetration into the bilayer makes part of the hydrophilic region invisible relative to the bulk D$_2$O (Kiselev et al. 2004). Therefore, the case of $\rho(x) =$ const gives underestimated value of $d = 36.70 \pm 0.02$ Å for DMPC vesicles at $T = 30^{\circ}$C (Zemlyanaya et al. 2005). Calculations, based on the SF approximation of $\rho(x)$, result in more reasonable value $d = 46.4 \pm 0.03$ Å and an additional introduction of the linear water distribution function in hydrophilic region of DMPC increases the membrane thickness $d = 47.4 \pm 0.04$ Å (Zemlyanaya et al. 2005). Generation of $\rho(x)$ based on the Gaussian functions gives $d = 50.6 \pm 0.8$ Å (Kiselev et al. 2004).

Water distribution function across the bilayer has sigmoidal form (Armen et al. 1998, Kiselev et al, 2004). In liquid phase, the contribution from the D$_2$O distribution function to the integrated neutron scattering length density of DMPC is sufficiently larger than the one from the polar head group (Kiselev et al. 2004). Competition between contributions of D$_2$O and polar head groups to the integrated neutron scattering length density allows one to simplify the approximation of $\rho(x)$ and decrease the number of fit parameters.

In present work, the SFF model is used to analyze the structure of the polydispersed population of DMPC vesicles in gel, liquid, and crystalline phases as studied by SANS. Two types of $\rho(x)$ functions are used for the evaluation of bilayer parameters, which depend on the bilayer hydration. Parameters of the vesicle population (<R>, σ for spherical shape or <a>, ε, σ for elliptical shape) and of bilayer (*d*, *D*, $n_w$) are calculated for the case of vesicles prepared via extrusion through 500 Å and 1000 Å diameter pores. The presented methods show what we can learn about vesicle structure from the SANS experiment on other types of lipid vesicles and vesicular based drug delivery systems.

**Materials and methods**

**Sample preparation**

DMPC was a gift from Lipoid (Moscow, Russia). $D_2O$ (99.9% deuteration) was purchased from Isotop (St. Petersburg, Russia). Samples for measurements were prepared by conventional extrusion technique. Heavy water and DMPC were mixed in a plastic tube and the tube was sealed. The DMPC concentration in the sample was 15mM (about 1 wt. %). The tube content was heated to a temperature above the main phase transition temperature and then cooled down to about − 20 °C. The cooling-heating cycle, accompanied by sample shaking, was repeated four times. From the dispersion of multilamellar vesicles thus obtained, extruded unilamellar vesicles were prepared in a single - step procedure according to (MacDonald et al. 1991) using the LiposoFast Basic extruder (Avestin, Ottawa, Canada). The multilamellar vesicle populations were prepared by extrusion through one polycarbonate filter (Nucleopore, Plesanton, USA) with pores of 500 Å (namely 500 Å extruded vesicles) or 1000 Å (namely 1000 Å extruded vesicles) diameter, mounted in the extruder fitted with two gas-tight Hamilton syringes (Hamilton, Reno, USA). The sample was subjected to 25 passes through the filter at a temperature higher than the main phase transition temperature of the DMPC. An odd number of passes were performed to avoid contamination of the sample by large and multilamellar vesicles, which might not have passed through the filter. The sample was filled into a quartz cuvette (Hellma, Müllheim, Germany) with a 2 mm sample thickness.

**SANS measurements**

SANS spectra were collected from unilamellar vesicles as function of temperature in the temperature interval 10 – 60°C in the range of scattering vector *q* from 0.02 Å$^{-1}$ to 0.15 Å$^{-1}$ (Ostanevich 1988) at YuMO small-angle time of flight spectrometer of Frank Laboratory of Neutron Physics (JINR, Dubna, Russia). Spectra were also obtained at *T* = 10°C, 20°C, and 30°C at SANS-1 spectrometer of the Swiss Spallation Neutron Source at the Paul Scherrer Institute (PSI), Switzerland. Three sample-to-detector distances 2 m, 6 m, and 20 m were used to obtain the SANS data over a wide *q* range from 0.0033 Å$^{-1}$ to 0.56 Å$^{-1}$. Neutron wavelength was 4.7 Å.

**Evaluation of vesicle parameters from SANS curve**

Simplest method of membrane thickness characterization is Guinier approximation of scattering curve (Feigin and Svergun 1987; Knoll et al. 1981; Gordeliy et al. 1993; Balgavy et al. 1998). The macroscopic cross section of vesicle population with *R* > *d* can be presented as (Gutberlet et al. 2000)

$$\frac{d\Sigma}{d\Omega}(q) = n \cdot \frac{d\Sigma}{d\Omega}(0) \cdot q^{-2} \cdot \exp(-R^2_t \cdot q^2), \qquad (3)$$

where *n* is a number of vesicles per unit volume, and

$$R_t^2 = \frac{d_G^2}{12}. \qquad (4)$$

Radius of gyration $R_t$ is determined from Guinier plot ($\frac{d\Sigma}{d\Omega}(q) \cdot q^2$ vs. $q^2$) and membrane thickness parameter $d_G$ is calculated from the Equation (4) (Gordeliy et al. 1993; Balgavy et al. 1998). The vesicle radius and polydispersity can not be determined in the Guinier approximation. The Equation (4) is valid for the case of large value of contrast, when scattering length density of $D_2O$ is larger than the average scattering length density of the bilayer. It is true for the case of "dry" bilayer, when penetration of $D_2O$ molecules inside the bilayer is negligibly small (Ibel and Stuhrmann 1975, Gordeliy et al. 1993). For DMPC vesicles in $D_2O$, the penetration of water molecules into the hydrophilic part of the bilayer influences the scattering length density distribution (Kiselev et al. 2004, Zemlyanaya et al. 2005). That is why $d_G$ is smaller relative to the real membrane thickness $d$ (Balgavy et al. 1998, Kiselev et al. 2001). Values of $d_G$ and $d$ are different on some constant $\Delta d_H$ depending on a membrane type and hydration

$$d = d_G + \Delta d_H. \tag{5}$$

A model of separated form factors (SFF) was proposed for an evaluation of the vesicle radius, the polydispersity and the internal membrane structure (Kiselev et al. 2002). The coherent macroscopic cross section of the monodispersed population of vesicles is defined by the formula

$$\frac{d\Sigma}{d\Omega}_{mon}(q) = n \cdot A^2(q) \cdot S(q) + IB, \tag{6}$$

where $n$ is the number of vesicles per unit volume, $A(q)$ is the scattering amplitude of a vesicle, $IB$ is the incoherent background, and $S(q)$ is the vesicle structure factor (Feigin and Svergun 1987, Kiselev et al. 2003a) For spherical unilamellar vesicle with radius $R$ (Kiselev et al. 2002, Zemlyanaya et al. 2005)

$$A(q) = 4\pi \cdot \int_{-d/2}^{d/2} \rho_c(x) \cdot \frac{\sin[(R+x) \cdot q]}{(R+x) \cdot q} \cdot (R+x)^2 \cdot dx, \tag{7}$$

where $\rho_c(x) = \rho(x) - \rho_{D2O}$ is contrast, the difference between scattering length densities of the bilayer $\rho(x)$ and the heavy water $\rho_{D2O}$. Integration of the Equation (7) gives exact expression for scattering amplitude of vesicle with separated parameters $R$, $d$, $\rho(x)$

$$A_{ves}(q) = 4\pi \cdot \frac{R^2}{qR} \cdot \sin(qR) \cdot \int_{-d/2}^{d/2} \rho_c(x) \cdot \cos(qx) \cdot dx$$
$$+ 4\pi \cdot \frac{R}{qR} \cdot \cos(qR) \cdot \int_{-d/2}^{d/2} \rho_c(x) \cdot x \cdot \sin(qx) \cdot dx \tag{8}$$

In the case of $R \gg d/2$, $R+x \approx R$, one can neglect the second term with respect to the one in Equation (8) (Kiselev et al. 2002, Zemlyanaya et al. 2005). The first term in Equation (8) presents the SFF model of the SANS scattering for vesicles. The macroscopic cross section of monodispersed population of vesicles in the frame of SFF model is written as

$$\frac{d\Sigma}{d\Omega}_{mon}(q) = n \cdot F_s(q,R) \cdot F_b(q,d) \cdot S(q) + IB, \tag{9}$$

where $F_s(q,R)$ is a form-factor of the infinitely thin sphere with radius $R$

$$F_s(q,R) = \left(4\pi \cdot \frac{R^2}{qR} \cdot \sin(qR)\right)^2, \tag{10}$$

$F_b(q,d)$ is a form-factor of the symmetric lipid bilayer.

$$F_b(q,d) = \left(\int_{-d/2}^{d/2} \rho_c(x) \cdot \cos(qx) \cdot dx\right)^2. \tag{11}$$

SFF model allows characterization of deformations of the vesicle shape from spherical to elliptical. In this case, instead of $F_S(q,R)$, the form factor of infinitely thin ellipse $F_E(q, a)$ is written as

$$F_E(q,a) = \int_0^1 A_E^2\left(q \cdot a \cdot \sqrt{1 + x^2 \cdot (\varepsilon^2 - 1)}\right) dx, \tag{12}$$

where $\varepsilon$ is an ellipse eccentricity, $a$ is a minor semi-axis, and the function $A_E(z) = 4\pi \cdot \varepsilon \cdot a^2 \cdot \frac{\sin(z)}{z}$.

The polydispersity of vesicle population is described by nonsymmetrical Schulz distribution (Hallet et al. 1991, Schmiedel et al. 2001)

$$G(R,<R>) = \frac{R^m}{m!} \cdot \left(\frac{m+1}{<R>}\right)^{m+1} \cdot \exp\left[-\frac{(m+1) \cdot R}{<R>}\right], \quad (13)$$

where $<R>$ is an average vesicle radius and $m$ is a coefficient of polydispersity. Relative standard deviation of the vesicle radius is given by

$$\sigma = \sqrt{\frac{1}{(m+1)}}. \quad (14)$$

Thus, the coherent macroscopic cross section of polydispersed vesicle population $I_{theor}(q,<R>,d)$ is calculated as:

$$I_{theor}(q,<R>,d) = \frac{\int_{R\min}^{R\max} \frac{d\Sigma}{d\Omega}_{mon}(q,R,d) \cdot G(R,<R>) \cdot dR}{\int_{R\min}^{R\max} G(R,<R>) \cdot dR}, \quad (15)$$

where $R_{min}$ and $R_{max}$ depend on the diameter of a polycarbonate filter.

The experimentally measured macroscopic cross section $I_{exp}(q)$ does not fully equal to theoretically calculated value of the coherent macroscopic cross-section $I_{theor}(q,<R>,d)$ due to the incoherent scattering background $IB$ from a sample and spectrometer resolution distortions. The experimentally measured macroscopic cross-section $\frac{d\Sigma}{d\Omega}(q)$ is approximated with good accuracy as

$$\frac{d\Sigma}{d\Omega}(q) = I_{theor}(q,<R>,d) + \frac{1}{2} \cdot \Delta^2 \cdot \frac{d^2 I_{theor}(q,<R>,d)}{dq^2} + IB, \quad (16)$$

for the case of $\frac{\Delta}{q} \leq 0.2$, where $\Delta^2$ is a second moment of a spectrometer resolution function (Ostanevich 1988).

Equation (16) was used to fit the SANS data by the $\chi^2$ minimizing package DFUMIL from the JINRLIB library (Silin 1967). The codes for fitting were developed in (Kiselev et al. 2004, Zemlyanaya et al. 2005). Value of incoherent background $IB$, number of vesicles per unit volume $n$, coefficient of polydispersity $m$, and thickness of the lipid bilayer $d$ are fitting parameters. There are some others too: average vesicle radius $<R>$ for spherical vesicles or average value of minor semi-axis $<a>$ along with eccentricity $\varepsilon$ for elliptical vesicles, as well as parameters of function $\rho(x)$, depicting the neutron scattering length density of the bilayer.

**Results and discussion**

**Dependence of the membrane thickness on temperature, Guinier approximation**

Figure 1 presents the dependence of the DMPC membrane thickness parameter $d_G$ on temperature. DMPC vesicles were prepared by extrusion through pores with diameter 1000 Å. SANS spectra were collected at YuMO spectrometer. Values of $d_G$ were evaluated in the Guinier plot via Equations (3) and (4). Membrane thickness parameter $d_G$ is found to be $44.2 \pm 0.8$ Å at $T = 10$ °C and $43.4 \pm 0.8$ Å at $T = 20$ °C. The value of $d_G$ of $38.8 \pm 0.8$ Å at $T = 30$ °C was calculated via data extrapolation in Figure 1. These values of $d_G$ underestimates the membrane thickness $d$ relative to the data from X-ray diffraction experiment which are 48.2 Å at $T = 10$ °C and 44.2 Å at $T = 30$ °C (Nagle and Tristram-Nagle 2000, Tristram-Nagle et al. 2002). The value of $\Delta d_H$ in Equation (5) is equal to 4 Å in the gel and 5.4 Å in the liquid phase of DMPC, which reflects larger DMPC hydration in the liquid phase. Guinier approximation describes the relative changes in the membrane thickness under temperature alteration rather well. Figure 1 demonstrates the decrease of the membrane thickness at the main phase transition temperature of 23ºC. The membrane thickness decreases on the value of 5.4 ± 1.6 Å on heating from 10 ºC to 30 ºC in an agreement with 4 Å decrease obtained from X-ray experiment.

**DMPC vesicles in the liquid crystalline phase, the HH approximation of $\rho(x)$**

The hydrophobic-hydrophilic (HH) approximation of the internal bilayer structure is used as $\rho(x)$ function in the liquid phase of DMPC (Schmiedel et al. 2004). It is based on the fact that the scattering length density of D$_2$O molecules in the hydrophilic region of bilayer is sufficiently larger then the scattering length density of polar head groups in liquid phase of DMPC (Kiselev et al. 2004). HH approximation of $\rho(x)$ is presented in Figure 2. Lipid bilayer consists of the two parts: hydrophobic and hydrophilic. Linear approximation is used for the water distribution function in the hydrophilic region, where $\rho_{D2O} = 6.4 \cdot 10^{10}$ cm$^{-2}$ and $\rho_{CH} = -0.36 \cdot 10^{10}$ cm$^{-2}$ are fixed parameters (Schmiedel et al. 2001, Zemlyanaya et al. 2005). The parameters of the DMPC bilayer determined from the SANS curve are the membrane thickness $d$ and the thickness of hydrophobic part $D$.

The number of water molecules $N_W$ and the surface area per DMPC molecule $A$ are determined from system

$$\begin{cases} A \cdot \dfrac{d}{2} = V_{DMPC} + N_W \cdot V_{D2O} & (17) \\ A \cdot \left( \rho_{CH} \cdot \dfrac{D}{2} + \dfrac{\rho_{D2O} - \rho_{CH}}{2} \cdot \dfrac{d-D}{2} \right) = l_{DMPC} + N_W \cdot l_{D2O} & (18) \end{cases}$$

where $V_{DMPC} = 1101$ Å$^3$ and $V_{D2O} = 30$ Å$^3$ are molecular volumes of DMPC and D$_2$O, $l_{DMPC} = 3.07 \cdot 10^{-12}$ cm and $l_{D2O} = 1.92 \cdot 10^{-12}$ cm are scattering lengths of the DMPC and D$_2$O molecules, respectively. Let us denote $l' = \dfrac{l_{DMPC}}{l_{D2O}}$ and $V' = \dfrac{V_{DMPC}}{V_{D2O}}$. One can define the average scattering length density of hydrated DMPC as

$$\bar{\rho} = \dfrac{D}{d} \cdot \rho_{CH} + \dfrac{d-D}{d} \cdot \dfrac{\rho_{D2O} - \rho_{CH}}{2} \tag{19}$$

The solution of Equations (17) and (18) is

$$N_W = \dfrac{\bar{\rho} \cdot V' - \rho_{D2O} \cdot l'}{\rho_{D2O} - \bar{\rho}}, \qquad A = 2 \cdot \dfrac{V_{DMPC} + N_W \cdot V_{D2O}}{d} \tag{20}$$

Figure 4 shows the experimentally measured and fitted SANS curves for the DMPC unilamellar vesicles at $T = 30$ ºC. Unilamellar vesicles were prepared by extrusion through pores with diameter of 500 Å. Experimentally measured macroscopic cross section was calculated by two different methods: by application of the exact expression for scattering amplitude using Equation (8) and by application of the SFF model using Equation (9). This allows one to check a validity of the SFF model and estimate the contribution of the second term in Equation (8). The obtained parameters are given in Table 1. Parameters of vesicle populations and internal membrane structure evaluated by exact expression and by SFF model for scattering amplitude coincide exactly. Consequently, the second term in the Equation 8 is negligibly small for vesicles with $<R> \geq 250$ Å.

For 500 Å extruded vesicles, the obtained value of $<R> = 275.6 \pm 0.5$ Å is in the agreement with radius of the pores (250 Å) used during vesicle extrusion. Membrane thickness $d$, evaluated on the base of the *HH* approximation of $\rho(x)$, is equal to 47.8±0.2 Å. Whereas, thickness of the hydrophobic region $D$ is 20.5±0.3 Å and of membrane hydrophilic part - $D_H = \dfrac{d-D}{2} = 13.7 \pm 0.5$ Å. The latter is larger than the one of the polar head groups $D_{PH} = 9$ Å, evaluated from X-ray diffraction (Nagle and Tristram-Nagle 2000), or $D_{PH}$ = 8.1±1.7 Å, evaluated from SANS (Kiselev et al. 2004). This result shows that water molecules penetrate into the hydrocarbon chain region on the 4.7 Å. The value of membrane thickness $d$ = 47.8±0.2 Å is 3.6 Å larger than $d$ = 44.2 Å evaluated from X-ray diffraction experiment on the giant multilamellar vesicles (Nagle and Tristram-Nagle 2000). The number of water molecules $Nw$ = 11.9±0.3 and surface area $A$ = 61.0±0.4 Å$^2$ of the DMPC molecule differ from $Nw$ = 7.2 and $A$ = 59.6 Å$^2$ for the giant multilamellar vesicles (Nagle and Tristram-Nagle 2000). Hydration and membrane thickness of the curved DMPC bilayer are larger relative to the flat bilayer.

The form of vesicles in the liquid phase cannot be ideally spherical. Probably, the vesicle shape fluctuates near the spherical form. These fluctuations serve as origins of the unilamellar vesicle deformation and orientation in the strong magnetic field. Unilamellar vesicles' deformation in the strong magnetic field 4T has been detected via SANS (Kiselev, unpublished). The elliptical form of the vesicle shape can be taken into account on the base of Equation (12). Table 2 presents the results for the case of 500 Å extruded vesicles, assuming that their form is elliptical. The value of eccentricity is 1.1, which demonstrates that vesicle shape is near to the spherical. Nevertheless, the membrane thickness has a value of 48.9±0.2 Å at ε=1.1 in comparison with $d$ = 47.8±0.2 Å for ε=0.

Important membrane parameter is the thickness of the polar head group $d_{PH}$. The information about $d_{PH}$ is hard to be obtained in the frame of the HH approximation. Large value of scattering length density of D$_2$O molecules screens the contribution from scattering length density of the polar head groups (Kiselev et al. 2004). Fitting procedure with $\rho(x)$ as sum of linear and step functions gives $d$ = 47.4±0.2 Å and $D$ = 17.3±0.5 Å (Zemlyanaya et al. 2005). The number of fitting parameters could be reduced by constrained value of $d_{PH}$ = 9 Å (Kucerka et al. 2004). This approach gives $d$ = 44.5±0.3 Å, $Nw$ = 6.8±0.3, and $A$ = 58.9±0.8 Å$^2$ at fitting of SANS curve in the range of $q$ from 0.03 Å$^{-1}$ to 0.14 Å$^{-1}$. An accuracy of the membrane thickness evaluation depends on a possibility to collect SANS curve in the region of $q$ corresponding to the bilayer form factor $F_b(q,d)$. Roughly, the first minimum of $F_b(q,d)$ corresponds to $q_m = \frac{2\pi}{d}$, and for $d$ = 45 Å $q_m$ = 0.14 Å$^{-1}$. Evaluation of the bilayer parameters could be improved by the measurements of SANS curve with maximum value of scattering vector $q_{max}$ about 0.3 Å$^{-1}$ as seen from the insert in Fig. 3. Data acquisition in the $q$ range from 0.30 Å$^{-1}$ to 0.56 Å$^{-1}$ is necessary for the evaluation of the incoherent background $IB$. Determination of $IB$, as fit parameter, improves an accuracy of the membrane thickness evaluation from SANS curve (Schmiedel et al. 2001). The enlargement of $q_{max}$ from 0.2 Å$^{-1}$ to 0.56 Å$^{-1}$ changes the value of $d$ from 42.5±0.3 Å to 47.4±0.2 Å and the value of $D$ from 11.0±0.9 to 17.3±0.5 Å (Zemlyanaya et al. 2005).

Membrane thickness $d$ = 45.5±0.6 Å is found for the unilamellar vesicles prepared by extrusion through pores with diameter 1000 Å. This value of membrane thickness is in the better agreement with $d$ = 44.2 Å obtained for giant multilamellar vesicles. Vesicles prepared via extrusion through pores 500 Å and 1000 Å are different in the membrane curvature and the polydispersity. The polydispersity (relative standard deviation of radius) increases from 27% to 47% on increasing of vesicle radius as seen from Fig. 4. Evaluated value of the average vesicle radius <R>=314.6±0.7 is sufficiently smaller than the radius of the polycarbonate pores (500 Å) and should be considered as an artifact. Dynamic light scattering (DLS) and freeze-fracture electron microscopy results show that average vesicle radius after extrusion is about the radius of the pores (MacDonald et al. 1991, Hallet et al. 1991, Patty and Frisken, 2003). The underestimation of the vesicle radius for 1000Å extruded vesicles has different reasons. First reason is a big value of the vesicles polydispersity (48%) relative to the one of 500 Å extruded vesicles (27%). Second one is an accuracy of the correction for spectrometer resolution. According to equation (10), positioning of first minimum in the form factor of infinitely thin sphere depends on the vesicle radius as $q_R = \frac{\pi}{R}$. Thus, $q_{500}$ = 0.0126 Å$^{-1}$ for vesicles with 500 Å diameter and $q_{1000}$ = 0.0063 Å$^{-1}$ for those with 1000 Å. Figure 5 demonstrates similar shift of $q_R$ for two types of polydispersed vesicle populations of 500 Å and 1000 Å extruded vesicles. The relative standard deviation of the spectrometer resolution function $\frac{\Delta}{q}$ = 0.53, 0.27, and 0.14 for $q$ = 0.0033 Å$^{-1}$, 0.0063 Å$^{-1}$, and 0.0126 Å$^{-1}$, respectively (Pedersen et al. 1990). For 1000 Å extruded vesicles, the accuracy of the Equation (16) decreases in the $q$ region, which is important for the evaluation of vesicles radius. Value of $q_{1000}$ = 0.0063 Å$^{-1}$ is not so far from the value of $q_{min}$ = 0.0033 Å$^{-1}$, scattering curve has only five experimental points between these two values. Measurements of the scattering curve with $q_{min}$ << 0.0033 Å$^{-1}$ and better spectrometer resolution are necessary for the correct evaluation of the average vesicle radius ≥ 500Å. Third possible reason for the underestimation of 1000Å extruded vesicle radius can be in a difference between the distribution statistics of 500Å and 1000Å extruded vesicles. Nonsymmetrical Schulz distribution well describes the 500Å extruded vesicles population. Weibull or some other nonsymmetrical distribution could be more suitable for the characterization of the vesicles population prepared via extrusion through 1000Å pores (Korgel et al. 1998). In any case, careful study of vesicle form and distribution requires the more powerful SANS instrument as discussed above for 1000 Å extruded vesicles.

Decreasing of bilayer thickness at diminishing of the membrane curvature reduces the number of water molecules $Nw$ from 11.9±0.3 to 10.8±0.3 and increases the surface area $A$ from 61.0±0.4 Å$^2$ to 62.6±1.0 Å$^2$, whereas, the hydrophobic thickness $D$ stays unchanged under the membrane curvature alteration. The structure and hydration of curved DMPC bilayer in the liquid phase depend on the membrane curvature. The bilayer thickness and hydration are altered at increasing of vesicle radius to its values for the flat bilayer.

It is known that unilamellar DMPC vesicles do not comprise a time-stable system. (Kiselev et al. 2003b). The membrane thickness and the hydration undergo alteration during system equilibration from the non-stable unilamellar vesicles population to the static state of the multilamellar one. Probably, the hydration of the DMPC bilayer is driving force in the transformation of the unilamellar vesicles to the multilamellar ones.

**DMPC vesicles in the liquid crystalline phase, the SF approximation of $\rho(x)$**

SF approximation of scattering length density across the lipid bilayer $\rho(x)$ is shown in Figure 3. The thickness of polar head group $d_{PH}$ = 9 Å was fixed as in the calculations done earlier (Kucerka et al. 2004). Vesicle shape is considered as spherical. The difference, relative to calculations in (Kucerka et al. 2004), is expressed in free value of the scattering length density in the region of polar head group, $\rho_{PH}$. The fitting parameters of the lipid bilayer are $d$ and $\rho_{PH}$. This allows one to make a calculation of water molecules located in the polar head group region $N_{w,\ PH}$ and the surface area $A$, taking into account that for SF approximation the average scattering length density of hydrated DMPC is defined as

$$\overline{\rho} = \frac{d - 2D_{PH}}{d} \cdot \rho_{CH} + \frac{2D_{PH}}{d} \cdot \rho_{PH} \qquad (21)$$

The calculated parameters are given in Table 3. Inside of the SF approximation, the membrane thickness $d$ = 45.5±0.7 Å does not show dependence on the membrane curvature and is the same as membrane thickness calculated based on the HH approximation for 1000 Å extruded vesicles. Both obtained values of the membrane thickness are larger than those of the giant multilamellar vesicles (44.2 Å). Other fitted and calculated parameters for 500 Å and 1000 Å extruded vesicles are the same within the experimental errors. Two calculated values of $N_{w,\ PH}$ = 6.8±0.6 and $N_{w,\ PH}$ = 8.0±0.6 are in good agreement with data for giant multilamellar vesicles 7.2 (Nagle and Tristram-Nagle 2000). Finally, the average value $N_{w,\ PH}$ =7.4 ± 0.6 can characterize the number of water molecules in the polar head group of DMPC at 30$^o$C. Comparison of the hydration calculated by HH and SF approximations allows one to calculate the number of water molecules in the region of hydrocarbon chains as 4.5 ± 0.9 for 500 Å and 3.4 ± 0.9 for 1000 Å extruded vesicles. Probably, the decrease of membrane curvature reduces the hydration of the hydrocarbon chains. But more precise experiments and data are necessary to check it. Surface area $A$, calculated based on the SF approximation, has average value 58 ± 1 Å, which is a little bit smaller than the value of 59.6 Å for giant multilamellar vesicles (Nagle and Tristram-Nagle 2000).

**DMPC vesicles in the gel and ripple phases, the SF approximation of $\rho(x)$**

Figure 6 shows the SANS spectrum from 500Å extruded vesicles measured at $T = 10^\circ$C (gel phase of DMPC) and $T = 30^\circ$C (liquid phase of DMPC). Without consideration of the vesicles polydispersity and the instrument resolution, the first minimum of $F_s(q,d)$ corresponds to $q_R = \frac{\pi}{R}$. The vesicle radius $R = 275.6$ Å corresponds to the $q_R = 0.011$ Å$^{-1}$. Shift of the $q_R$ to the larger value of $q$ reflects the decrease of average vesicle radius at temperature lowering as seen from Figure 6.

The consideration of SANS curves in the Guinier region of $q$ from 0.03 Å$^{-1}$ to 0.15 Å$^{-1}$ shows the different exponents at $T = 10\,^\circ$C and $T = 30\,^\circ$C (equation 3). It is visually seen that at temperature decreasing from 30$^\circ$C to 10$^\circ$C the radius of gyration of the bilayer $R_t$ increases. This simple analysis corresponds to the dependence of the membrane thickness parameter $d_G$ on temperature as shown in Figure 1.

The number of water molecules in the bilayer depends on the lipid phase. For giant multilamellar DPPC vesicles, the number of water molecules in the region of polar head groups decrease from 8.6 in the liquid phase to 3.7 in the gel phase (Nagle and Tristram-Nagle 2000). SF approximation of $\rho(x)$ with fixed thickness of the polar head region $d_{PH}$ is more appropriate for gel and ripple phases of the DMPC bilayer (Figure 3). Value of $\rho_{CH} = -0.39 \cdot 10^{10}$ cm$^{-2}$ was used as fixed parameter in the calculations for gel and ripple phases of DMPC.

Measured SANS curves at $T = 10^\circ$C and $T = 20^\circ$C for 500 Å and 1000 Å extruded vesicles were fitted based on the SF approximation of $\rho(x)$. The deformation of spherical shape to elliptical was taken into account. Opposite the liquid phase, extruded 500 Å vesicles in gel and ripple phases have pronounced elliptical form with eccentricity $\varepsilon = 1.6$. 1000 Å extruded vesicles have $\varepsilon = 1.1$ and can be considered as spherical taking into account problems in the determination of vesicle radius. The internal membrane structure did not show the dependence on the eccentricity and membrane curvature for both vesicle populations. The results of fitting and calculation are given in Table 4 for extruded 500 Å vesicles in gel ($T = 10^\circ$C) and ripple ($T = 20^\circ$C) phases. Results for extruded 1000 Å vesicles are given in Table 5. The average values of the minor semi-axis $a = 185$ and the major semi-axis $b = 300$ at $T = 10^\circ$C change to the $a = 187$ and $b = 303$ at $T = 20^\circ$C for vesicles prepared by extrusion through pores with 500 Å diameter. These values are in the agreement with radius of pores. Similar to the liquid DMPC phase at 30$^\circ$C, the average radius of 1000 Å extruded vesicles is underestimated relative to the radius of pores and DLS results. Polydispersity of vesicle population has smaller values in gel and ripple phase relative to ones in the liquid phase. For 500 Å extruded vesicles, the polydispersity increases from the value of 21% in gel phase to the value of 26% in liquid phase. The average area of the 500 Å extruded vesicles $S = 4\pi\varepsilon a^2$ is equal to $9.9 \cdot 10^5$ Å$^2$ at $T = 30^\circ$C. Average area of the vesicle decreases to the values of $7.2 \cdot 10^5$ Å$^2$ in ripple and $7.0 \cdot 10^5$ Å$^2$ in gel phases. The decrease of vesicle area during the phase transitions from liquid to gel phase should increase the membrane thickness. Membrane thickness increases from $47.8 \pm 0.2$ Å at 30$^\circ$C to $49.1 \pm 0.7$ Å at 10$^\circ$C.

Unlike the liquid DMPC phase, the difference in the membrane curvature for 500 Å and 1000 Å extruded vesicles has no influence on the membrane thickness and hydration in gel and ripple phases. Values of the membrane thickness $d = 49.6 \pm 0.5$ Å and the surface area $A = 49.2 \pm 0.9$ Å$^2$ for 1000Å extruded vesicles are little bit larger than values $d = 48.2$ Å and $A = 47$ Å$^2$ determined for giant multilamellar DMPC vesicles in gel phase (Tristram-Nagle et al. 2002). Giant multilamellar DMPC vesicles can be considered as membrane with zero curvature. For the gel phase, numbers of water molecules in the region of polar head groups per DMPC molecule $N_{w, PH}$ obtained for 500 Å and 1000 Å extruded vesicles are the same. Average value of $N_{w,PH} = 5.6 \pm 0.5$ obtained in gel phase is smaller than hydration of liquid phase, $N_{w,PH} = 7.4 \pm 0.6$. This difference in the hydration of gel and liquid phases is sufficiently smaller than the corresponding difference in the hydration of DPPC.

Opposite to the liquid and gel phases of DMPC, information about internal membrane structure and hydration in ripple phase is hard to be evaluated through the X-ray diffraction experiment due to the limited number of the diffraction peaks. Parameters of the vesicle population and the bilayer structure of DMPC vesicles in ripple phase are given in Table 4 and Table 5. For 500Å extruded vesicles, the evaluated vesicle parameters are the same in gel and ripple phases within the range of experimental errors. For 1000 Å extruded vesicles, the membrane thickness decreases from the value of 49.6±0.5 Å in gel phase to 48.3±0.6 Å in ripple phase. Finally, one can use parameters evaluated for 1000 Å extruded vesicles $d = 48.3 \pm 0.6$ Å, $N_{w,PH} = 5.9 \pm 0.4$, and $A = 50.4 \pm 0.8$ Å to characterize DMPC bilayer structure in ripple phase.

**Guinier approximation for the flat membrane, connection with SFF model**

It is commonly believed that Guinier approximation describes the SANS curve at $q \to 0$. It is true for globular particles with one typical size (Guinier and Fournet 1955). Vesicles have two typical sizes: radius and membrane thickness, which are different in the scale of one - two orders by a magnitude. SANS curve from vesicles have two Guinier regions. First one is Guinier region of the vesicle size at $q \to 0$ and other one is Guinier region of the membrane. The sense and validity of the membrane Guinier approximation is clarified below.

Let one consider lipid membrane as the one with $\rho_c(x) \equiv \Delta\rho = $ const. This approximation conventionally named as homogenous approximation (Feigin and Svergun 1987). Really, lipid membrane is non-homogenous object, $\rho_c(x) \neq$ const. Homogenous approximation can describe only some part of the experimental scattering curve (Feigin and Svergun 1987). This part of the curve can be obtained via a comparison of experimental and calculated SANS curves (Kiselev et al. 2002). According to the SFF model, the scattering amplitude of vesicle for $\rho_c(x) \equiv \Delta\rho = $ const and $R >> d_G$ is written as (Kiselev et al. 2002)

$$A(q) = \left(4\pi R^2 \cdot \frac{\sin(qR)}{qR}\right) \cdot \left(\frac{2\Delta\rho}{q} \cdot \sin\left(\frac{qd_G}{2}\right)\right). \tag{22}$$

The approximation of the harmonic function by exponent

$$\frac{\sin x}{x} \approx e^{-\left(\frac{x}{\sqrt{6}}\right)^2} \tag{23}$$

is valid for $x < 1$. Equation (22) can be rewritten as

$$A(q) = \left(S \cdot \frac{Sin(qR)}{qR}\right) \cdot \Delta\rho \cdot d_G \cdot \exp\left(-\frac{(qd_G)^2}{24}\right), \tag{24}$$

where $S = 4\pi R^2$ is the vesicle surface area. It is important to note that transition from the harmonic function to the exponent has no request of $q \to 0$. The validity of this transformation is $q \cdot \frac{d_G}{2} < 1$. The exponential presentation of the scattering amplitude is the principal property of the Guinier approximation. For $q < \frac{2}{d_G}$, the vesicle form factor $F = A^2$ is written as

$$F(q) = \left(\frac{4\pi S \cdot \sin^2(qR)}{q^2}\right) \cdot (\Delta\rho \cdot d)^2 \cdot \exp\left(-\frac{(qd_G)^2}{12}\right). \tag{25}$$

Let us consider a flat lipid bilayer, which corresponds to $R \to \infty$. In this case, $\Delta q = \frac{\pi}{R} \to 0$, where $\Delta q$ is distance between minima of the $\sin^2(qR)$. Any SANS spectrometer has uncertainty in the values of measured $q$, as well as any vesicle population has uncertainty in the value of $R$. One can, therefore, use average value $\overline{\sin^2(qR)} = \frac{1}{2}$ instead of $\sin^2(qR)$. Finally, using Equation (4) for $R_t$, one obtains

$$F(q) = \left(\frac{2\pi S}{q^2}\right) \cdot (\Delta\rho \cdot d_G)^2 \cdot \exp(-q^2 \cdot R_t^2) \tag{26}$$

Equation (26) is Guinier approximation of the form factor for the flat bilayer with infinitely large area $S$ and membrane radius of gyration $R_t$. Form factor of the infinitely thin flat membrane has $1/q^2$ behavior. Form factor of the membrane with thickness parameter $d_G$ has exponential form and is characterized at $q \rightarrow 0$ by asymptotic properties of the function $F(q) \cdot q^2$.

Equation (26) was evaluated from SFF model on the base of two approximations: $\Delta\rho = $ const and $R \rightarrow \infty$. This equation limits a possibility to receive some information about internal membrane structure. Internal membrane structure is a case of $\Delta\rho \neq$ const and requires $q$ values to be around the region of Guinier. It can explain the difference between the values of DMPC membrane thickness for 500 Å extruded vesicles at $T = 30°C$ obtained via application of the Guinier approximation, $d = 44.5 \pm 0.3$ Å and presented in Tables 3 $d = 45.5 \pm 0.7$ Å (Kucerka et al. 2004).

Dependence of the membrane thickness parameter $d_G$ on temperature as shown in Figure 1 can be corrected now for the values of membrane thickness calculated for 1000 Å extruded vesicles at 10°C, 20°C, and 30°C. Obtained values of parameter $\Delta d_H$ are 5.4 Å, 4.9 Å, and 6.7 Å respectively for 10°C, 20°C, and 30°C. Membrane thickness $d$ is calculated using Equation (5) with an assumption that $\Delta d_H$ has permanent values in gel, ripple, and liquid phases. Figure 7 shows the obtained dependence of the DMPC membrane thickness on temperature for 1000 Å extruded vesicles. Main future in temperature dependence $d(T)$ is a sharp increase in the membrane thickness at the temperature of the main phase transition, $T = 23°C$. Obtained anomalous $d(T)$ behavior supports the critical fluctuations of membrane thickness rather than fluctuations of water layer at the temperature of main phase transitions (Zhang et al. 1995, Lemmich et al. 1995).

**Conclusions**

Separated Form Factor model of the small angle scattering can be used for data interpretation for vesicles with radius larger than 250 Å. Shultz distribution well describes the vesicle shape and size for vesicles prepared via extrusion through pores of 500 Å. Average vesicle shape is near to spherical (eccentricity 1.1 ) for liquid phase of DMPC and deviates sufficiently from sphere in gel and ripple phases (eccentricity 1.6). The average vesicle radius and the polydispersity can be evaluated correctly from SANS curve with minimum measured scattering vector $q_{min} = 0.0033$ Å$^{-1}$ for vesicles with radius about 250 Å. It should be emphasized that SANS experiment with $q_{min} < 0.0033$ Å$^{-1}$ is necessary for correct evaluation of radius $\geq 500$ Å. Approximation of the neutron scattering length density across DMPC bilayer in liquid phase as hydrophobic and hydrophilic regions with linear water distribution describes the internal bilayer structure without any preliminary structural information. Step Function approximation of the scattering length density allows evaluating of DMPC membrane thickness and neutron scattering length density of the polar head group in gel, ripple and liquid phases based on the preliminary structural information about thickness of the polar head group. It was shown that thickness of the bilayer depends on the membrane curvature. The DMPC bilayer thickness of the curved unilamellar vesicles ($49.6 \pm 0.5$ Å and $45.5 \pm 0.6$ Å in gel and liquid phases, respectively) is larger relative to that of multilamellar vesicles with nearly flat membrane (48.2 Å and 44.2 Å in gel and liquid phase, respectively). Interpretation of the SANS data is based on the calculation of the scattering length density across the membrane bilayer. Information about specific volume of the lipid molecule allows one to receive additional structural information from SANS. In our study, a preliminary knowledge of the DMPC molecular volume allows one to calculate bilayer hydration and lipid surface area. The X-ray diffraction has been successfully applied to characterize the internal membrane structure in gel and liquid phase. However, the diffraction normally does not give accurate information about membrane structure in ripple phase. This drawback of diffraction technique can be overcome by SANS. The internal membrane structure and hydration of ripple DMPC phase were evaluated through the SANS experiment in present article.

The results presented here emphasize the importance of appropriate models and approximations for the interpretation of SANS data. Usefulness of SANS technique for the characterization of the internal bilayer structure and hydration was shown for vesicular systems. Thus, presented herein methods can be used for the characterization of vesicular based drug delivery systems.

**Acknowledgements**

This work is partly based on the experiments performed at Swiss spallation neutron source SINQ, Paul Scherrer Institute, Villigen, Switzerland. The authors are grateful to Prof. Dr. V.L. Aksenov for support of this study, Prof. Dr. H. Schmiedel and Prof. Dr. A.M. Balagurov for fruitful discussions. The investigation was supported by the Ministry of Science and Technology RF (contract № 40.012.1.1.1148), Grant of Leading Scientific Scholl, and RFBR (grant №03-01-00657). The authors would like to thank Lipoid (Moscow) for the gift of DMPC.

**References**

Armen PS, Uitto OD, Feller SE (1998) Phospholipid component volumes: determination and application to bilayer structure calculations. Biophys. J. 75: 734-744

Balgavy P, Dubnickova M, Uhrikova D, Yaradaikin S, Kiselev M, Gordeliy V (1998) Bilayer thickness in unilamellar extruded egg yolk phosphatidylcholine liposomes: a small-angle neutron scattering study. Acta Physica Slovaca 48: 509-533

Balgavy P, Dubnickova M, Kucerka N, Kiselev MA, Yaradaikin SP, Uhrikova D (2001) Bilayer thickness and lipid interface area in unilamellar extruded 1,2-diacylphosphatidylcholine liposomes: A small-angle neutron scattering study. Biochim Biophys Acta 1521: 40-52

Cevc G, Schatzlein A, Richardsen H (2002) Ultradiformable lipid vesicles can penetrate the skin and other semi-permiable barriers unfragmented. Evidence from double label SLSM experiments and direct size measurements. Biochim Biophys Acta 1564: 21-30

Feigin LA, Svergun DI (1987) Structure analysis by small-angle X-ray and neutron scattering. Plenum Publishing Corporation, New York

Glatter O (1977) Data evaluation in small angle scattering: calculation of the radial electron density distribution by means of indirect fourier transformation. Acta Physica Austriaca 47: 83-102

Glatter O (1980) Evaluation of small-angle scattering data from lamellar and cylindrical particles by the indirect transformation method. J Appl Cryst 13: 577-584

Gordeliy VI, Golubchikova LV, Kuklin AI, Syrykh AG, Watts A (1993) The study of single biological and model membranes via small angle neutron scattering. Progr Colloid Polym Sci 92: 252-257.

Gordeliy VI, Kiselev MA (1995) Definition of lipid membrane structural parameters from neutronographic experiments with the help of the strip function model. Biophys J 69: 1424-1428

Guinier A, Fournet G (1955) Small-angle scattering of X-rays. New York, John Wiley & Sons, Inc.

Gutberlet T, Kiselev M, Heerklotz H, Klose G (2000) SANS study of mixed POPC/$C_{12}E_n$ aggregates. Physica B 381-383: 276-278

Hallet FR, Watton J, Krygsman P (1991) Vesicle sizing. Number distributions by dynamic light scattering. Biophys. J. 59: 357-362

Jin AJ, Huster D, Gawrisch K, Nossal R (1999) Light scattering characterization of extruded lipid vesicles. Eur Biophys J 28: 187-199

Kiselev MA, Lesieur P, Kisselev AM, Lombardo D, Killany M, Lesieur S (2001). Sucrose solutions as prospective medium to study the vesicle structure: SAXS and SANS study. J Alloys Compounds 328: 71-76

Kiselev MA, Lesieur P, Kisselev AM, Lombardo D, Aksenov VL (2002) Model of separated form factors for unilamellar vesicles. J Applied Physics A 74: S1654-S1656

Kiselev MA, Lombardo D, Kisselev AM, Lesieur P, Aksenov VL (2003a) Structure factor of dimyristoylphosphatidylcholine unilamellar vesicles: small-angle X-ray scattering study. Surface 11: 20-24 (in russian)

Kiselev MA, Wartewig S, Janich M, Lesieur P, Kiselev AM, Ollivon M, Neubert R (2003b). Does sucrose influence the properties of DMPC vesicles. Chem Phys Lipids 123: 31-44

Kiselev MA, Zemlyanaya EV, Aswal VK (2004) SANS study of unilamellar DMPC vesicles: Fluctuation model of a lipid bilayer. Crystallography Reports 49: s136-s141

Kiselev MA, Ryabova NY, Balagurov S. Dante S, Hauss T, Zbytovska J, Wartewig S, Neubert RHH (2005) New insights into structure and hydration of stratum corneum lipid model membrane by neutron diffraction. . Europ Biophys accepted

Knoll W, Haas J, Stuhrmann H, Fuldner HH, Vogel H, Sackmann E (1981) Small-angle neutron scattering of aqueous dispersions of lipids and lipid mixtures. A contrast variation study. J Appl Cryst 14: 191-202

Korgel B, van Zanten JH, Monbouquette HG (1998) Vesicle size distributions measured by flow field-flow fractionation. Biophys J 74: 3264-3272.

Kucerka N, Kiselev MA, Balgavy P (2004) Determination of bilayer thickness and lipid surface area in unilamellar dimyristoylphosphatidylcholine vesicles from small-angle neutron scattering curves: a comparison of evaluation methods. Eur Biophys J 33: 328-334

Lemmich J, Mortensen K, Ipsen JH, Honger T, Bauer R, Mouritsen OG (1995) Pseudocritical behaviour and unbinding of phospholipids bilayers. Phys Rev Let 75: 3958-3961

MacDonald RC, MacDonald RI, Menco BP, Takeshita K, Subbarao NK, Hu LR (1991) Small-volume extrusion apparatus for preparation of large, unilamellar vesicles. Biochim Biophys Acta 1061: 297-303

Nagayasu A, Uchiyama K, Kiwada H (1999) The size of liposomes: a factor which affects their targeting efficiency to tumors and therapeutics activity of liposomal antitumor drugs. Adv Drug Delivery Rev 40: 75-87

Nagle JF, Tristram-Nagle S (2000) Structure of lipid bilayers. Biochim Biophys Acta 1469: 159-195

Ostanevich YuM (1988) Time-of-flight small-angle scattering spectrometers on pulsed neutron sources. Macromol. Chem. Macromol. Symp. 15: 91-103

Patty PJ, Frisken BJ (2003) The pressure-dependence of the size of extruded vesicles. Biophys. J. 85: 996 – 1004.

Pedersen JS (1997) Studies of soft condenced matter by SANS, SAXS and reflectometry. 5[th] Summer School on Neutron Scattering, Lyceum Alpinum, Zuoz, Switzerland

Pedersen JS, Posselt D, Mortensen K (1990) Analytical treatment of the resolution function for small-angle scattering. J Appl Cryst 23: 321-333

Pencer J, Hallet R (2000): Small-angle neutron scattering from large unilamellar vesicles: An improved method for membrane thickness determination. Phys Rev E 61: 3003-3008

Pencer J, White GF, Hallet FR (2001) Osmotically induced shape changes of large unilamellar vesicles measured by dynamic light scattering. Biophys J 81: 2716-2728

Schmiedel H, Joerchel P, Kiselev M, Klose G (2001) Determination of structural parameters and hydration of unilamellar POPC/C12E4 vesicles at high water excess from neutron scattering curves using a novel method of evaluation. J Phys Chem B 105: 111-117

Schmiedel H, Almasy L, Wang R, Klose G (2004) Multilamellarity, structure and hydration of extruded POPC vesicles by SANS. *In* Condensed Matter Physics with Neutrons at the IBR-2 Pulsed Reactor. E14-2004-148, JINR, Dubna

Silin IN (1967) Standard minimization code on the base of the least square method. JINR Communication 11-3362, Dubna

Tristram-Nagle S, Liu Y, Legleiter J, Nagle JF (2002) Structure of gel phase DMPC determined by X-ray diffraction. Biophys J 83: 3324-3335

Wiener MC, White SH (1991) Fluid bilayer structure determination by the combined use of X-ray and neutron diffraction. Biophys J 59: 162-173

Zhang R, Sun W, Tristram-Nagle S, Headrick RL, Suter RM, Nagle JF (1995) Critical fluctuations in membranes. Phys Rev Let 74: 2832-2835

Zemlyanaya EV, Kiselev MA (2002) Determination of the unilamellar dimyristoylphosphatidylcholine vesicle structure from the small-angle scattering data. JINR Communication P3-2002-163, Dubna

Zemlyanaya EV, Kiselev MA, Aswal VK (2005) Structure of the unilamellar dimyristoylphosphatidylcholane vesicles. A small-angle neutron scattering study. J Computational Methods in Applied Sciences and Engineering: in press, ArXiv: physics/0411029

**Table 1** Results for DMPC vesicles in liquid phase ($T=30^{\circ}$C) based on the HH approximation of $\rho(x)$. $D_F$ is diameter of the pores used at extrusion, $<R>$ is average vesicle radius, $\sigma$ is vesicle polydispersity, $d$ is membrane thickness, $D$ is thickness of the hydrophobic core, $N_w$ and $A$ are number of water molecules and surface area per DMPC molecule, $IB$ is value of incoherent background.

| Model | $D_F$, Å | $<R>$, Å | $\sigma$, % | $d$, Å | $D$, Å | $N_w$ | $A$, Å$^2$ | $IB$, cm$^{-1}$ |
|---|---|---|---|---|---|---|---|---|
| SFF | 500 | 275.6±0.5 | 27 | 47.8±0.2 | 20.5±0.3 | 11.9±0.3 | 61.0±0.4 | 0.007 |
| Exact | 500 | 275.7±0.4 | 27 | 47.8±0.2 | 20.5±0.3 | 11.9±0.3 | 61.0±0.4 | 0.007 |
| SFF | 1000 | 314.6±0.7* | 48 | 45.5±0.6 | 20.8±0.4 | 10.8±0.4 | 62.6±1.0 | 0.007 |

\* $<R>$=450 Å from DLS (Hallet et al. 1991)

**Table 2** Results for DMPC vesicles in liquid phase ($T=30^{\circ}$C) based on the HH approximation of $\rho(x)$ and elliptical deformations. $D_F$ is diameter of the pores used at extrusion, $<a>$ is average value of minor semi-axis, $\varepsilon$ is eccentricity, $\sigma$ is vesicle polydispersity, $d$ is membrane thickness, $D$ is thickness of the hydrophobic core, $N_w$ and $A$ are number of water molecules and surface area per DMPC molecule, $IB$ is value of incoherent background. SFF model.

| $D_F$, Å | $<a>$, Å | $\varepsilon$ | $\sigma$, % | $D$, Å | $D$, Å | $N_w$ | $A$, Å$^2$ | $IB$, cm$^{-1}$ |
|---|---|---|---|---|---|---|---|---|
| 500 | 266±2 | 1.11±0.02 | 26 | 48.9±0.2 | 19.9±0.4 | 12.8±0.3 | 60.7±0.5 | 0.007 |

**Table 3** Results for DMPC vesicles in liquid phase ($T=30^{\circ}$C) based on the SF approximation of ρ(x). $D_F$ is diameter of the pores used at extrusion, $<R>$ is average vesicle radius, $\sigma$ is vesicle polydispersity, $d$ is membrane thickness, $\rho_{PH}$ is scattering length density in the region of polar head group, $N_{w, PH}$ and $A$ are number of water molecules in region of polar head group and surface area per DMPC molecule, $IB$ is value of incoherent background. SFF model.

| $D_F$, Å | $<R>$, Å | $\sigma$, % | $d$, Å | $\rho_{PH}$, $10^{10}$ cm$^{-2}$ | $N_{w, PH}$ | $A$, Å$^2$ | $IB$, cm$^{-1}$ |
|---|---|---|---|---|---|---|---|
| 500 | 275.1±0.5 | 27 | 45.5±0.7 | 3.7±0.2 | 6.8±0.6 | 57±1 | 0.007 |
| 1000 | 296±2* | 48 | 45.7±0.7 | 4.0±0.2 | 8.0±0.6 | 59±1 | 0.007 |

\* $<R>$=450 Å from DLS (Hallet et al. 1991)

**Table 4** Results for DMPC 500 Å extruded vesicles in gel ($T=10^{o}$C) and ripple ($T=20^{o}$C) phases based on the SF approximation of $\rho(x)$. $T$ is temperature, $<a>$ is average value of minor semi-axis, $\varepsilon$ is eccentricity, $\sigma$ is vesicle polydispersity, $d$ is membrane thickness, $\rho_{PH}$ is scattering length density of polar head group, $N_{w,\ PH}$ and $A$ are number of water molecules in region of polar head group and surface area per DMPC molecule, $IB$ is value of incoherent background. SFF model.

| $T$, $^{o}$C | $<a>$, Å | $\varepsilon$ | $\sigma$, % | $d$, Å | $\rho_{PH}$, $10^{10}$ cm$^{-2}$ | $N_{w,\ PH}$ | $A$, Å$^2$ | $IB$, cm$^{-1}$ |
|---|---|---|---|---|---|---|---|---|
| 10 | 185±1 | 1.62 | 21 | 49.1±0.7 | 3.7±0.2 | 5.2±0.5 | 48.8±0.9 | 0.005 |
| 20 | 187±1 | 1.63 | 22 | 47.9±0.7 | 3.6±0.3 | 5.3±0.5 | 50±1 | 0.006 |

**Table 5** Results for DMPC 1000 Å extruded vesicles in gel ($T=10^{o}$C) and ripple ($T=20^{o}$C) phases based on the SF approximation of $\rho(x)$. $T$ is temperature, $<R>$ is average vesicle radius, $\sigma$ is vesicle polydispersity, $d$ is membrane thickness, $\rho_{PH}$ is scattering length density of polar head group, $N_{w,\ PH}$ and $A$ are number of water molecules in region of polar head group and surface area per DMPC molecule, $IB$ is value of incoherent background. SFF model.

| $T$, $^{o}$C | $<R>$, Å | $\sigma$, % | $d$, Å | $\rho_{PH}$, $10^{10}$ cm$^{-2}$ | $N_{w,\ PH}$ | $A$, Å$^2$ | $IB$, cm$^{-1}$ |
|---|---|---|---|---|---|---|---|
| 10 | 309.0±0.4* | 37 | 49.6±0.5 | 3.6±0.2 | 5.9±0.5 | 49.2±0.9 | 0.006 |
| 20 | 316±1* | 35 | 48.3±0.6 | 3.8±0.2 | 5.9±0.4 | 50.4±0.8 | 0.006 |

* $<R>$=450 Å from DLS (Hallet et al. 1991)

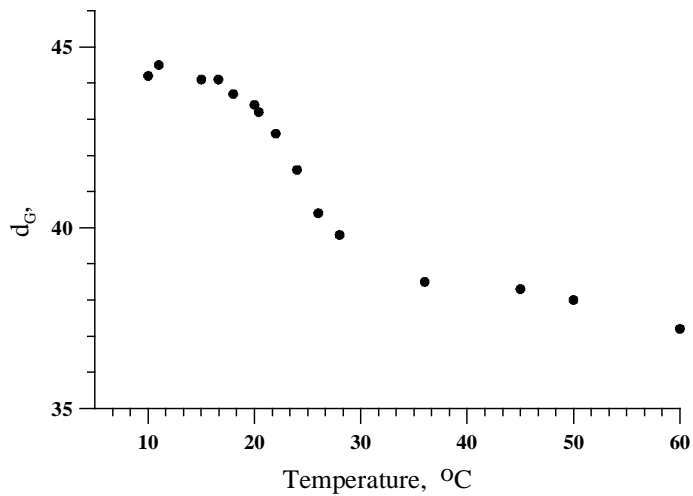

Fig. 1. Dependence of DMPC membrane thickness parameter $d_G$ on temperature for the 1000 Å extruded vesicles. Values of $d_G$ were evaluated from SANS curves in the Guinier approximation.

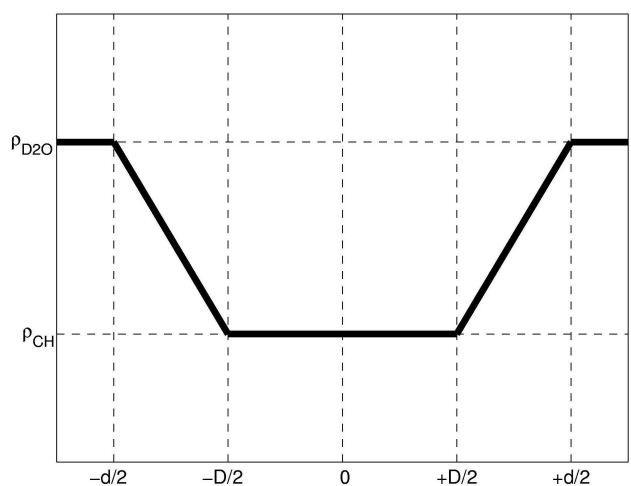

Fig. 2. Hydrophobic – Hydrophilic (HH) approximation of the scattering length density across the lipid bilayer $\rho(x)$. $d$ is the membrane thickness, and $D$ is the thickness of the hydrophobic part of the membrane.

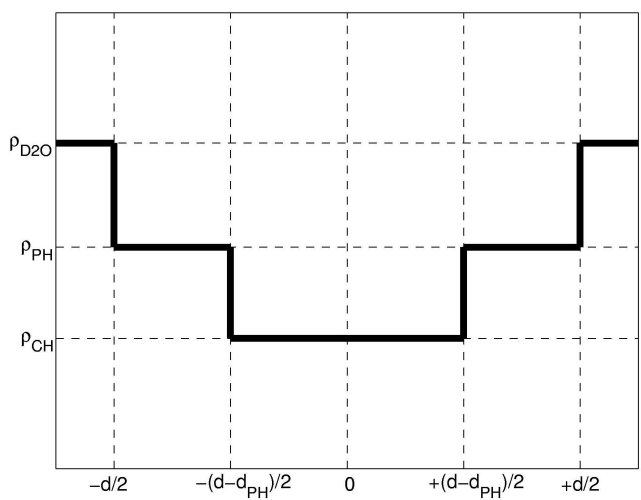

Fig. 3. Step Function (SF) approximation of the scattering length density across the lipid bilayer $\rho(x)$. $d$ is the membrane thickness, and $d_{PH}$ is the thickness of the polar head group. $\rho_{PH}$ is scattering length density in the region of the polar head group.

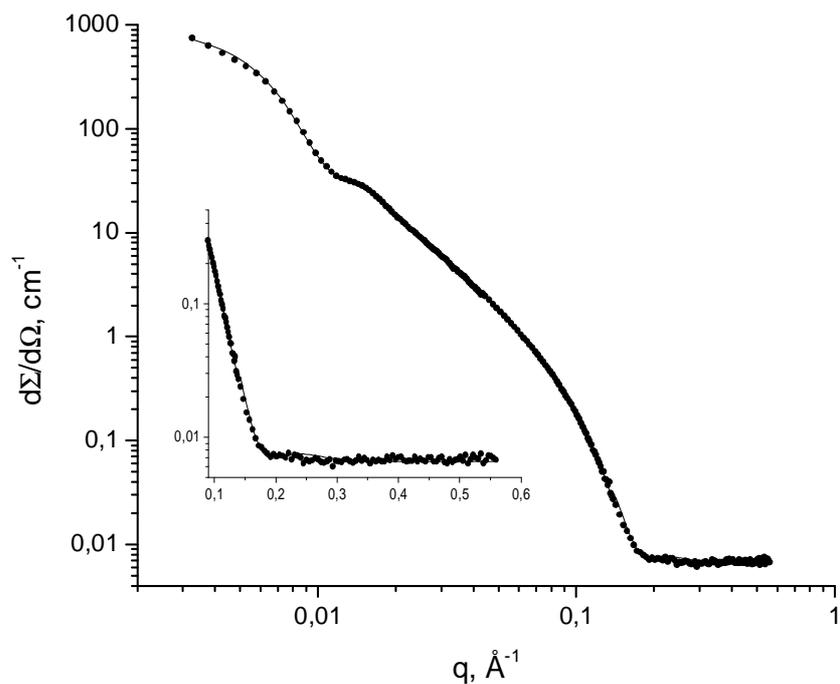

Fig. 4. Experimental macroscopic cross-section of the unilamellar vesicles population at $T = 30^{\circ}$C (dots) for vesicles extruded through pores of 500 Å diameter and fitting curve (solid line). The inset shows the magnified curve for large $q$.

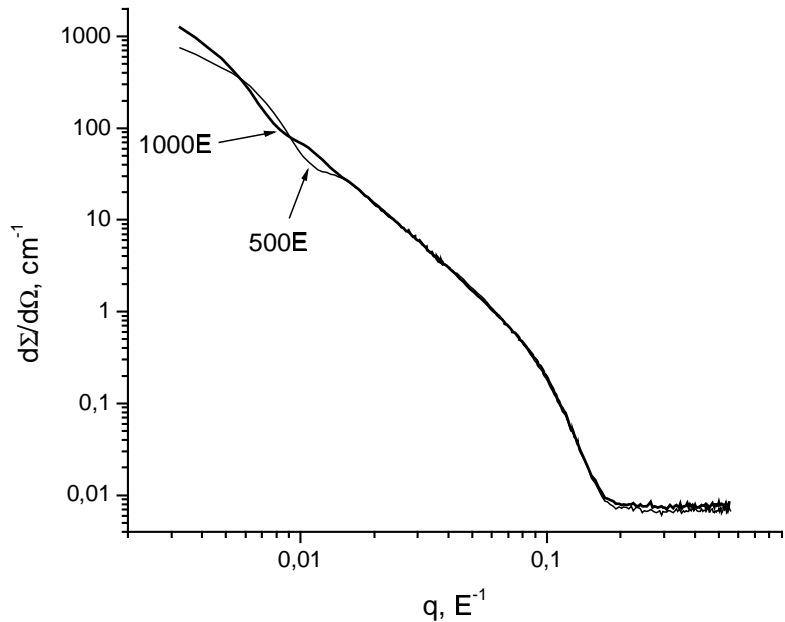

Fig. 5. Experimental macroscopic cross-sections of the unilamellar vesicles populations at $T = 30^{\circ}$C for vesicles prepared by extrusion through pores of 500 Å and 1000 Å diameters.

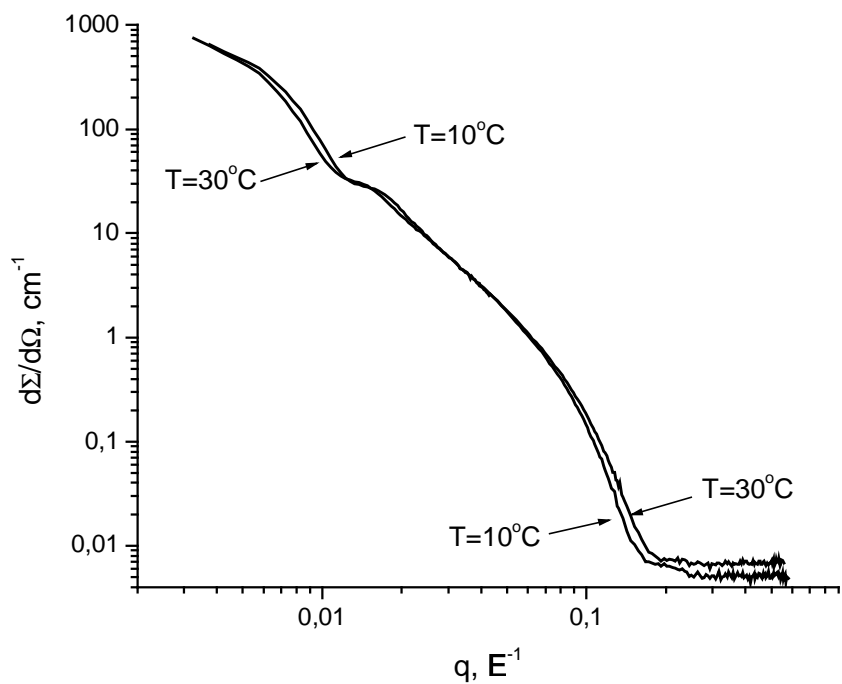

Fig. 6. Experimental macroscopic cross-sections of the unilamellar vesicles populations at $T = 30^{\circ}$C and $T = 10^{\circ}$C for vesicles prepared by extrusion through pores of 500 Å diameter.

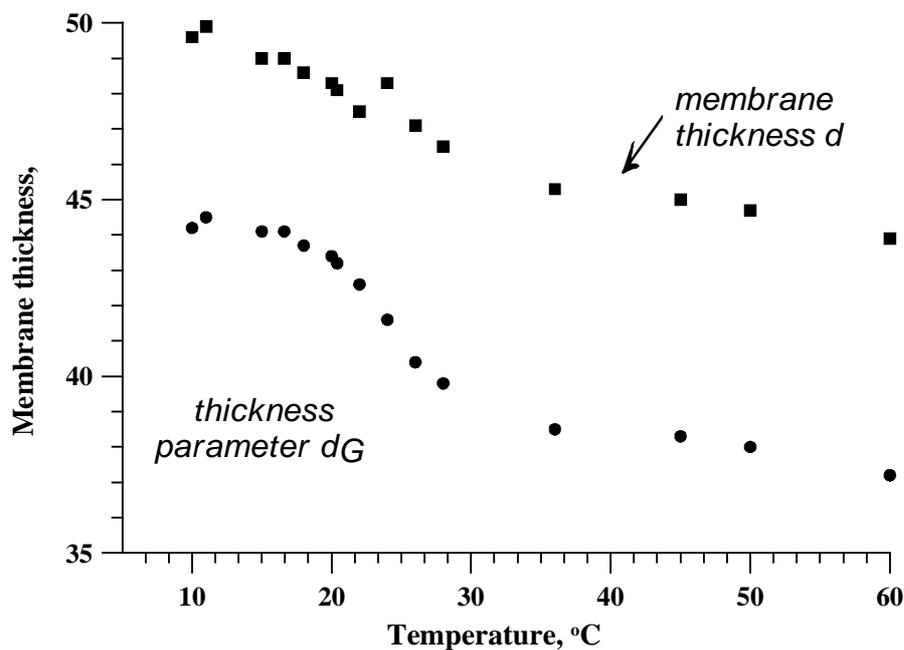

Fig. 7. Dependence of DMPC membrane thickness parameter $d_G$ and membrane thickness $d$ on temperature for 1000 Å extruded vesicles.